\documentclass[12pt]{iopart}

\usepackage{graphicx}
\usepackage[german,english]{babel}
\usepackage{iopams}
\usepackage{subfigure}
\usepackage{braket}

\begin{document}

\title[DFT++ method implemented with projector augmented waves]{General DFT++ method implemented with projector augmented waves: Electronic structure of SrVO$_3$ and the Mott Transition in Ca$_{2-x}$Sr$_{x}$RuO$_4$}

\author{M. Karolak, T. O. Wehling, F. Lechermann, A. I. Lichtenstein}
\address{I. Institut f{\"u}r Theoretische Physik, Universit{\"a}t Hamburg, Jungiusstra{\ss}e 9, D-20355 Hamburg, Germany}

\ead{mkarolak@physik.uni-hamburg.de}

\begin{abstract}

The realistic description of correlated electron systems has taken an important step forward a few years ago as the combination of
density functional methods and the dynamical mean-field theory was conceived. This framework allows access to both high and low energy physics and is capable of the description of the specific physics of strongly correlated materials, like the Mott metal-insulator transition. A very important step in the procedure is the interface between the band structure method and the dynamical mean-field theory and its impurity solver. We present a general interface between a projector augmented wave based density functional code and many-body methods based on Wannier functions obtained from a projection on local orbitals. The implementation is very flexible and allows for various applications. Quantities like the momentum resolved spectral function are accessible. We present applications to SrVO$_3$ and the metal-insulator transition in Ca$_{2-x}$Sr$_{x}$RuO$_4$.                   
                                                                                                                                                                                                                                      
\end{abstract}


\maketitle

\section{Introduction}
\label{introduction}

The theoretical description of strongly correlated electron systems is one of the main challenges in contemporary condensed matter physics. Since the intricate properties of such systems have become accessible experimentally, a unified theory that can account for material-specific quantities as well as complicated many-body behaviour is needed. Traditionally, the state of the art electronic structure theory that captures material specific physics is density functional theory (DFT) (see e.g Ref. \cite{jones_gunnarsson} for a review). This theory can be applied to itinerant systems (e.g. metals) where the kinetic energy dominates localization, but it fails for the strongly correlated ones (transition metal oxides, Kondo systems, etc.). The incorrect behaviour of the DFT for strongly correlated systems can be remedied by augmenting the DFT with a local Hubbard-like interaction. The resulting lattice Fermion model with \textit{ab initio} parameters from DFT is then solved by the dynamical mean-field theory (DMFT) (see e.g. Ref. \cite{dmft_review} for a review). The resulting methodologies are commonly referred to as LDA+DMFT \cite{Anisimov_Poteryaev,Lichtenstein_Katsnelson} (see Ref. \cite{ldadmft_review} for a recent review) or in the static Hartree-Fock limit LDA+U \cite{ldapu91,ldapu_review}. In such acronyms LDA is used synonymously with DFT, since the methods are not specifically limited to the LDA functional. In general a method interfacing DFT with an additional many-body treatment can be called DFT++ following Ref. \cite{Lichtenstein_Katsnelson}. DFT++ combines the ab initio capabilities of DFT to include material specific effects (crystal structures, etc.)
with the capabilities of model Hamiltonians to account for the physics of strongly correlated electrons (Mott transitions, Kondo physics, etc.).
From a practical point of view, all DFT++ methods base
on a Hubbard type interaction being added to the Kohn-Sham Hamiltonian \cite{kohn_sham}, which
serves as bare, ``non-interacting" starting point. \\
An important point at the interface between DFT and many-body methods is a suitable choice for a basis of the one-electron Kohn-Sham states. The local Hubbard-like interaction in the correlated subspace of orbitals (usually $d$ or $f$ orbitals) is best represented in a set of localized orbitals. Thus, earliest implementations of the DFT+DMFT used the linear-muffin-tin-orbital (LMTO) basis and represented the correlated subspace with the subset of LMTO's with the specific character. This choice is certainly sensible, however, it might not be optimal and other basis sets have been investigated. A basis set, that has been heavily used in recent years, is the basis of Wannier functions. These have been utilized in different flavours in the context of DFT+DMFT: Anisimov \etal \cite{anisimov_wannier}, Amadon \etal \cite{Amadon08}, Haule \etal \cite{haule2010} and Aichhorn \etal \cite{Aichhorn2009} used different schemes for projections onto a subset of Bloch functions, Pavarini \etal \cite{pavarini_andersen} used Nth order muffin-tin orbitals (NMTOs) and Lechermann \etal \cite{lechermann_wannier} used maximally localized Wannier functions (MLWFs) \cite{marzani_vanderbilt}.\\
In this paper we lay out methodological details concerning the use of projections of Bloch states onto local orbitals, as introduced in Refs. \cite{Amadon08,korotin_wannier} in the framework of the projector augmented wave method \cite{Bloechl_PAW}. We will refer to the method in what follows as PLO (projected local orbitals), as in Ref. \cite{Amadon08}. The purpose of this paper is to present an implementation of a general DFT++ method in the PAW basis set using the VASP \cite{kresse_hafner,kresse_joubert} code and to explore different methods to construct the underlying Wannier functions. We compare the present implementation with maximally localized Wannier Functions produced in different basis sets and NMTO calculations.

The paper is organized as follows: In Section \ref{method} we present a general outline of the approach, with special attention to specific details concerning the PAW method. The methodological details will be illustrated using SrVO$_3$ as a test system. As a more challenging application of the DFT++ approach a brief discussion of the metal-insulator transition in Ca$_{2-x}$Sr$_{x}$RuO$_4$ is included.

\section{DFT++ in the PAW framework}
\label{method}
\subsection{General Presentation}
\label{general}

\begin{table}
\caption{\label{hops} Intersite hopping integrals $t^{xyz}_{yz,yz}$ for SrVO$_3$ from PLO compared with maximally localized Wannier functions in different basis sets and Nth order muffin-tin orbitals. Energies in meV. (\footnotemark[1] from Ref. \cite{lechermann_wannier})}
\begin{indented}
\item[]\begin{tabular}{@{\extracolsep{\fill}} lccccccccc}
\br
$xyz$ & 001 & 100 & 011 & 101 & 111 & 002 & 200 \\
\mr
PLO(VASP)           & -261.5 & -28.5 & -87.1&  6.7   & -6.4 & 7.5 & 0.1     \\
MLWF(VASP+WANNIER90)& -261.5 & -28.5 & -87.2&  6.7  & -6.4 & 7.5 & 0.1    \\
MLWF(FLAPW)\footnote[1]{}         & -266.8 & -29.2 & -87.6& 6.4   & -6.1   & 8.3  & 0.1\\
NMTO(LMTO-ASA)\footnotemark[1]      & -264.6 & -27.2 & -84.4& 7.3   & -7.6   & 12.9  & 3.5\\
\br
\end{tabular}
\end{indented}
\end{table}

The \textit{ab initio} treatment of correlated electron systems requires the calculation of Green functions and hybridization functions in terms of local orbitals. This is readily achieved when using a basis set, which is localized in real space, such as linear muffin-tin orbitals or Gaussian basis sets. However, many implementations of the density functional theory use a delocalized plane wave basis set. This has the advantages, that the basis set is simple, universal and its convergence is controlled in principle by a single parameter, the energy cutoff. The projector augmented wave method (PAW), being a representative of plane-wave based methods, is a fast and accurate way of implementing DFT \cite{Bloechl_PAW}. It formally is a so-called all electron method meaning it takes into account all electrons in the problem. At the heart of the method is the following theorem, stating that the wave function $\ket{\Psi}$ can be decomposed exactly as follows
\begin{eqnarray}\ket{\Psi}&=&\mathcal{T}\ket{\tilde{\Psi}}=\Bigl(1+\sum_{i}\tau_i\Bigr)\ket{\tilde{\Psi}}\nonumber\\
&=&\ket{\tilde{\Psi}}+\sum_{i}\left(\ket{\phi_i}-\ket{\tilde{\phi}_i}\right)\braket{\tilde{p}_i|\tilde{\Psi}}.
\label{eqn:PAW_decomp}
\end{eqnarray}
Here $\mathcal{T}$ is the operator denoting the transformation, which is different from the identity only inside an augmentation region, $\ket{\tilde{\Psi}}$ is the so called \textit{pseudo} wave function and the sum runs over all augmentation channels $i$. 
As in the construction of pseudopotentials, physical partial waves $\ket{\phi}$ are solutions of the Schr\"odinger equation of isolated atoms, while the corresponding auxiliary (pseudo) functions $\ket{\tilde{\phi}}$ are chosen to match $\ket{\phi}$ outside the
augmentation spheres, being smooth inside and continuously differentiable in all
space. The projectors $\ket{\tilde{p}_i}$ are finally defined by 
\begin{equation*}
\braket{\tilde{p}_i|\tilde{\phi}_j}=\delta_{ij},
\end{equation*}
where the tilde as usually in the PAW formalism discriminates \textit{pseudo} from \textit{physical} quantities.
The Kohn-Sham equations in PAW representation are obtained by applying
the variational principle to the total energy functional with respect to the auxiliary wave functions: Since the transformation operator $\mathcal{T}$ does not depend on the
electron density, the Kohn-Sham equations transform according to 
\[\mathcal{T}^\dagger H_{\mathrm{KS}} \mathcal{T} \ket{\tilde{\Psi}_\mathbf{k}}=\varepsilon_\mathbf{k} \mathcal{T}^\dagger \mathcal{T} \ket{\tilde{\Psi}_\mathbf{k}}\]
Here, $H_{\mathrm{KS}}$ is the Kohn-Sham Hamiltonian, so
that above equation is a Schr\"odinger type equation, but with the overlap operator 
occurring on the right hand side. To solve the equation the auxiliary wave functions
are expanded in terms of plane waves:
\[\tilde{\Psi}_\mathbf{k}(\mathbf{r})=\braket{\mathbf{r}|\tilde{\Psi}_\mathbf{k}}=\sum_{\mathbf{G}}c_{\mathbf{k},\mathbf{G}}\exp\left(i(\mathbf{k}+\mathbf{G})\mathbf{r}\right).\]

Following Ref. \cite{lechermann_wannier}, the desired quantity for an implementation of a DFT++ method is a projection \\ 
$\mathcal P^\mathcal C=\sum_L \ket{L}\bra{L}$ of the full Kohn-Sham Green function $G_{\mathrm{KS}}(\omega)$ on a set of localized orbitals $\{\ket{L}\}$:
\begin{equation}
G^{\mathcal C}_{\mathrm{KS}}(\omega)=\mathcal P^\mathcal C G_{\mathrm{KS}}(\omega)\mathcal P^\mathcal C.
\label{eqn:Gloc_appx}
\end{equation}
The subspace $\mathcal C={\rm span}(\{\ket{L}\})$ is usually termed correlated subspace. It is the subspace of orbitals in which many-body correlations play a major role and where corrections to the DFT will be considered.
In plane-wave based calculations, $G_{\rm KS}(\omega)$ is available in terms of a (truncated) set of Bloch states $\ket{K}$ that are eigenstates of the Kohn-Sham Hamiltonian $H_{\rm KS}\ket{K}=\epsilon_K\ket{K}$:
\begin{equation}
\label{eqn:GKS_appx}
 G_{\rm KS}(\omega)=\sum_K\frac{\ket{K}\bra{K}}{\omega+i0^+-\epsilon_K}.
\end{equation}
Inserting equation (\ref{eqn:GKS_appx}) into equation (\ref{eqn:Gloc_appx}) shows that one needs to evaluate projections of the type $\braket{L|K}$ in order to access the matrix elements $( G^\mathcal C_{\rm KS}(\rm \omega))_{LL'}$ of the local Green function.
In most cases the correlated orbitals are $d$ or $f$ orbitals, which are to a good approximation localized inside the PAW augmentation spheres. For $\ket{L}$ within these spheres and given the PAW decomposition of a Bloch state $\ket{K}$ one obtains
\begin{equation*}
\braket{L|K}=\braket{L|\tilde K}+\bra{L}\left(\sum_{i}(\ket{\phi_i}-\ket{\tilde\phi_i})\braket{\tilde{p}_i|\tilde K}\right),
\end{equation*} 
which simplifies for a converged partial wave expansion to 
\[\braket{L|K}=\sum_{i}\braket{L|\phi_i}\braket{\tilde{p}_i|\tilde K}.\] 
The index $i$ of the augmentation functions $\ket{\phi_i}$ includes site $\mu$, angular momentum $l$ and $m$ as well as an index $\nu$ labeling the radial function: $i=(\mu,l,m,\nu)$.
In practice, the localized orbitals can be also chosen to be angular momentum eigenstates at a given site $\mu$, which leads to
\begin{equation}
\label{eqn:LK_proj_full}
 \braket{L|K}=\sum_\nu \braket{L|\phi_\nu}\braket{\tilde{p}_\nu|\tilde K}
\end{equation}
with $\nu$ abbreviating $i=(\mu,l,m,\nu)$ where $\mu,\,l,$ and $m$ are fixed. In the PAW approach, the first augmentation function, $\nu=0$, for each channel is usually taken to be an atomic eigenfunction  (c.f. Ref. \cite{Bloechl_PAW}). Defining the correlated subspace in terms of atomic eigenfunctions leads consequently to $\ket{L}=\ket{\phi_{\nu=0}}$:
\begin{equation}
\label{eqn:LK_proj_phi}
 \braket{\phi_{\nu=0}|K}=\sum_\nu \braket{\phi_{\nu=0}|\phi_\nu}\braket{\tilde{p}_\nu|\tilde K}.
\end{equation} 
As higher augmentation functions, $\nu>0$, are in general not orthogonal to the $\ket{\phi_{\nu=0}}$ state, evaluation of equation (\ref{eqn:LK_proj_full}), requires accounting for the overlaps of the form $\braket{\phi_{\nu=0}|\phi_{\nu'}}$. This approach has been implemented to the AB-INIT code \cite{abinit1,abinit2} by Amadon \cite{Amadon08}. In the present work we have used the Vienna \textit{Ab-Initio} Simulation Package (VASP) \cite{kresse_hafner,kresse_joubert} to implement the same method. It will be referred to as PLO(A) in what follows.

The first order approximation to equation (\ref{eqn:LK_proj_phi}) has also been implemented to investigate how accurate it is. In this case one \textit{only} uses the first term of the sum in equation (\ref{eqn:LK_proj_phi})
\begin{equation}
\label{eqn:LK_proj_zero}
 \braket{\phi_{\nu=0}|K}=\braket{\phi_{\nu=0}|\phi_{\nu=0}}\braket{\tilde{p}_{\nu=0}|\tilde K}
\end{equation}
while disregarding all other terms. This approximation will be referred to as PLO(0). In addition to the above methods we have implemented another scheme: As in the LDA+U scheme implemented in VASP \cite{VASP_LDApU,PAW+U} we choose
\begin{equation}
\label{eqn:LK_proj_approx_abs}
 |\braket{L|K}|^2=\sum_{\nu,\nu'} \braket{\tilde K|\tilde{p}_\nu}\braket{\phi_{\nu}|\phi_{\nu'}}\braket{\tilde{p}_{\nu'}|\tilde K}.
\end{equation}
Hence, the absolute value of the projections is in this scheme given by the projection onto a subspace of augmentation channels with given angular momentum, $(l,m)$. The phase is determined by
\begin{equation}
\label{eqn:LK_proj_approx_phase}
 \arg\left(\braket{L|K}\right)=\arg\left(\sum_\nu\braket{\tilde{p}_\nu|\tilde K}\right).
\end{equation}
This particular construction will be referred to as PLO(V) in the following. If higher augmentation channels are negligible, $|\braket{\tilde{p}_{\nu=0}|\tilde K}|\gg|\braket{\tilde{p}_{\nu>0}|\tilde K}|$, and additionally $\braket{\phi_\nu|\phi_{\nu^\prime}}=\delta_{ij}$ (which is not the case in general) equation (\ref{eqn:LK_proj_zero}), equation (\ref{eqn:LK_proj_phi}) as well as equations (\ref{eqn:LK_proj_approx_abs})-(\ref{eqn:LK_proj_approx_phase}) become formally identical. The approach defined in equation (\ref{eqn:LK_proj_approx_abs})-(\ref{eqn:LK_proj_approx_phase}) differs from the approach in equation(\ref{eqn:LK_proj_full}) in one important point. In the former approach a specific radial function $\ket{\phi_0}$ is used for the projection and only overlaps of the type $\braket{\phi_0|\phi_\nu}$ are taken into account, whereas in the latter approach the radial dependence is averaged out by including general overlaps $\braket{\phi_\nu|\phi_{\nu^{\prime}}}$.
As constructed, the projections in  equation (\ref{eqn:LK_proj_phi}) as well as in equations (\ref{eqn:LK_proj_approx_abs}) and (\ref{eqn:LK_proj_approx_phase}) are not properly normalized for two reasons: (1) the Bloch basis is incomplete since only a limited number of Bloch bands is included and (2) the PAW augmentation functions are in general not orthonormal. In our implementation we orthonormalize the projection matrices by the following Wannier type construction: By definition, the localized states $\ket{L}$ are labeled by site and angular momentum indices: $L=(\mu,l,m)$. We split the site index $\mu=\bf R+T$ such that $\bf R$ labels the position within the unit cell and $\bf T$ is the Bravais lattice vector of the unit cell in which $\mu$ is located. This allows us to construct the Bloch transform of the localized states,
\begin{equation}
\ket{L_{\bf k}}=\sum_T e^{i\bf kT}\ket{L_{\bf T}}, \label{eqn:L_BlochTransform}
\end{equation}
where $\bf k$ is from the first Brillouin zone and $\ket{L_{\bf{T}}}\equiv\ket{L}=\ket{\mu,l,m}$ with $\mu=\bf{R+T}$. The sum in equation (\ref{eqn:L_BlochTransform}) runs over the Bravais lattice. Labeling the Bloch states $\ket{K}=\ket{\mathbf{k},n}$ by their crystal momentum, $\bf k$, and band index, $n$, we normalize our projection matrices $\mathcal P^\mathcal C_{Ln}(\mathbf{k})=\braket{L_\mathbf{k}|\mathbf{k},n}$ using the overlap operator
\begin{equation}
 O_{LL'}(\mathbf{k})=\sum_n \mathcal P^\mathcal C_{Ln}(\mathbf{k})\left(\mathcal P^\mathcal C_{L'n}(\mathbf{k})\right)^*
\end{equation}
in
\begin{equation}
 \bar{\mathcal P}^\mathcal C_{Ln}(\mathbf{k})=\sum_{L'}\left[O(\mathbf{k})\right]^{-1/2}_{LL'}P^\mathcal C_{L'n}(\mathbf{k}).
\end{equation}

These orthonormalized projection matrices are calculated once at the beginning of any calculation and can then be used to obtain the
local Green function of the correlated orbitals from the full Bloch Green function
\[G_{LL'}^{\mathcal{C}}(\omega)=\sum_{\mathbf{k},nn'}\bar{\mathcal P}^\mathcal C_{Ln}(\mathbf{k})G^{Bloch}_{nn'}(\mathbf{k},\omega)\left(\bar{\mathcal P}^\mathcal C_{L'n'}(\mathbf{k})\right)^\ast.\]

Similarly the hybridization function, $\Delta(\omega)$, is available. It is related to the local Green function by

\begin{equation}
G^{-1}(\omega)=\omega+i\delta-\epsilon_d-\Delta(\omega),
\label{eqn:Green_Hyb}
\end{equation}

where $\epsilon_d$ is the static crystal field. equation (\ref{eqn:Green_Hyb}) is a matrix equation with $G$, $\Delta$, and $\epsilon_d$ being $(\dim\,\mathcal C)\times (\dim\,\mathcal C)$ matrices, in general. To separate the hybridization from the static DFT crystal field, we numerically evaluate the limit $\omega\to \infty$, where $\omega-G^{-1}(\omega)\to \epsilon_d$.

In a DFT+DMFT calculation the projection matrices $\bar{\mathcal P}^\mathcal C_{Ln}(k)$ are used for up- and downfolding quantities like the Green function and the self energy in the course of the iterative DMFT procedure in exactly the same way as shown for the local Green function above. For example, the self energy obtained by an impurity solver for the effective impurity model $\Sigma^{\mathcal{C}}_{LL^\prime}
(\omega)$ can be upfolded to the Bloch basis as follows: 
\[\Sigma^{Bloch}_{nn^\prime}(\mathbf{k},\omega)=
\sum_{LL^\prime}\left(\overline{P}^{\mathcal C}_{Ln}(\mathbf{k})\right)^{\ast}~\Sigma^{\mathcal{C}}_{LL^\prime}
(\omega)~\overline{P}^{\mathcal C}_{L^\prime n^\prime}(\mathbf{k}).\] 
Since the self energy in DMFT is a purely local quantity, the index $\mathbf{k}$ on $\Sigma^{Bloch}_{nn^\prime}(\mathbf{k},\omega)$ reflects the momentum dependence brought about by the projection matrices. The presented projection scheme allows for the inclusion of both correlated and uncorrelated states in the procedure. Therefore, information about the interplay of correlated orbitals with their uncorrelated ligands can be obtained.

With these quantities available we can proceed to perform DFT++ calculations of different kinds. The local Green function or the hybridization function can be used as an input for impurity calculations, e.g. impurities on surfaces as discussed for the examples of Co on Cu(111) in Ref. \cite{co_on_cu} as well as Co on graphene in Ref. \cite{co_on_graphene}. The DFT+DMFT implementation based on the projectors from equations (\ref{eqn:LK_proj_approx_abs})-(\ref{eqn:LK_proj_approx_phase}) has been compared to NMTO \cite{pavarini_andersen,pavarini_review} and other plane wave Wannier function based implementations \cite{trimarchi,korotin_wannier} for different cases (e.g for Sr$_2$RuO$_4$, LaTiO$_3$, YTiO$_3$, NiO), where the results were in good accordance \cite{mk_unpub,ca2ruo4_prl}. 

\subsection{Many Body Methods, Impurity Solvers}

\begin{figure}[t]
 \centering
  \includegraphics[width=0.5\textwidth]{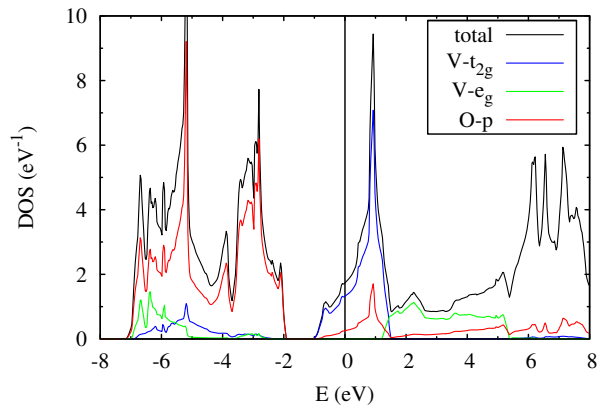}
\caption{DFT(LDA) total and orbital projected density of states for SrVO$_3$}
\label{lda_dos}
\end{figure}

The goal of the present DFT++ implementation is to interface DFT with a subsequent many-body treatment. Since our implementation yields Green functions and hybridization functions, we can apply it to any many-body method based on those quantities. 
A widely used DFT++ methodology is the above mentioned DFT+DMFT. Here the local downfolded DFT Green function, which contains the full hybridization function, serves as an initial input to the impurity solver. We have used here the Hirsch-Fye Quantum Monte Carlo \cite{hirsch_fye} (HF-QMC) solver. In all cases an additional input is the local Coulomb interaction matrix. The DFT+DMFT Hamiltonian can then be written as follows for a two or three band model
\begin{eqnarray} 
H=&&H_{\rm KS}+\frac{1}{2}\sum_{i,m, m^\prime,\sigma}U_{mm^{\prime}} n_{im,\sigma}n_{im^\prime,-\sigma} \label{hamiltonian} \\
&&+\frac{1}{2}\sum_{i,m\neq m^\prime,\sigma}(U_{mm^\prime}-J_{mm^\prime})n_{im,\sigma}n_{im^\prime,\sigma}\nonumber \\
\nonumber \\
&&+\frac{1}{2}\sum_{i,m\neq m^\prime,\sigma}J_{mm^\prime}\Bigl(c^{\dagger}_{im,\sigma}c^{\dagger}_{im^\prime,-\sigma}c_{im,-\sigma}c_{im^\prime,\sigma}\nonumber \\
&&\hspace*{70pt}+c^{\dagger}_{im,\sigma}c^{\dagger}_{im,-\sigma}c_{im^\prime,\sigma}c_{im^\prime,-\sigma}\Bigr).\nonumber
\end{eqnarray} 
Here $n^{\sigma}_{im}$ is the number operator for electrons at site $i$ with 
spin $\sigma$ in an orbital characterized by the quantum number $m$. The single-electron part given by  
$H_{\rm KS}$ contains all material specific quantities.

\begin{figure}[t]
 \centering
  \includegraphics[width=0.5\textwidth]{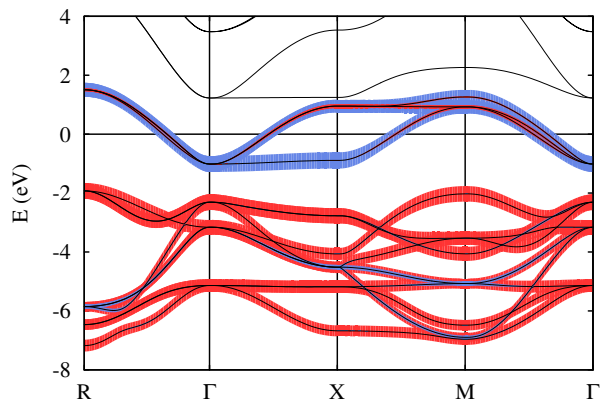}
\caption{DFT(LDA) band structure of SrVO$_3$ with fatbands highlighting the V $t_{2g}$ (blue) states and the O $2p$ states (red).}
\label{fat_bands}
\end{figure}

The terms in the first two lines of equation (\ref{hamiltonian}) contain only density-density type interactions. In the
following lines, however, the so-called spin-flip and pair-hopping terms appear in which two
electrons in different orbitals flip their spins or transfer to other orbitals. If the HF-QMC method is used as an impurity solver, one is limited to
density-density interactions. The recently developed continuous-time Monte Carlo (CT-QMC) \cite{rubtsov_ctqmc,werner_ctqmc}
impurity solvers can process the full rotationally invariant U-matrix. It is important to note that the neglect of the spin-flip and pair-hopping terms happens for
technical reasons in the first place. It is not always a reasonable assumption that the exchange
coupling J is small with respect to U and can be left out to a good approximation.
The neglected terms can be of importance in certain systems \cite{sakai06,co_on_cu,delta_pu,ca2ruo4_prl}. If the local part of the Hamiltonian (\ref{hamiltonian}) is solved by the QMC method continuous hybridization functions can be taken into account. While QMC is formally exact,
the method provides data with statistical noise on the imaginary axis, which makes analytic continuation problematic.

 \begin{figure}[t]
  \centering
   \includegraphics[width=0.5\textwidth]{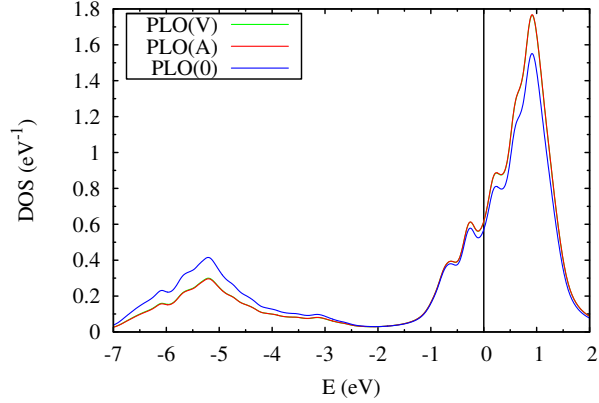}
 \caption{Effective impurity local density of states as obtained by PLO(V) (green), PLO(A) (red), and PLO(0) (blue) methods using the V $t_{2g}$ as well as the O $2p$ bands in the projection.}
 \label{wannier_bands_12}
 \end{figure}

\section{Applications}
\label{applications}
\subsection{SrVO$_3$}
\subsubsection{DFT}

The correlated metal SrVO$_3$ has become a common testing ground for first-principles many-body techniques and has been the subject of multiple theoretical and experimental investigations \cite{mit_review,fujimori1992,maiti2001,inoue1995,sekiyama2004,liebsch2003,nekrasov2005,
pavarini_andersen,yoshida2005,wadati2006,solovyev2006,nekrasov2006,eguchi2006,Aichhorn2009}. It has a perfectly cubic perovskite structure (space group Pm$\overline{3}$m) with a lattice constant of 3.84 \AA\ \cite{garcia-jaca}. The V ion is surrounded by O ligands in a perfectly octahedral configuration leading to a splitting of the $d$ orbitals into $t_{2g}$ and $e_g$ crystal-field states. The density of states (figure \ref{lda_dos}) and band-structure (figure \ref{fat_bands}) obtained using the VASP code and LDA reveals three isolated bands at the Fermi level which are formed by the degenerate $t_{2g}$ states of vanadium. The bandwidth of this block amounts to $2.5$ eV. The Van Hove singularity 1eV above the Fermi level in the DOS corresponding to these bands shows their predominantly two dimensional character. 

\begin{figure}[t]
 \centering
  \includegraphics[width=0.5\textwidth]{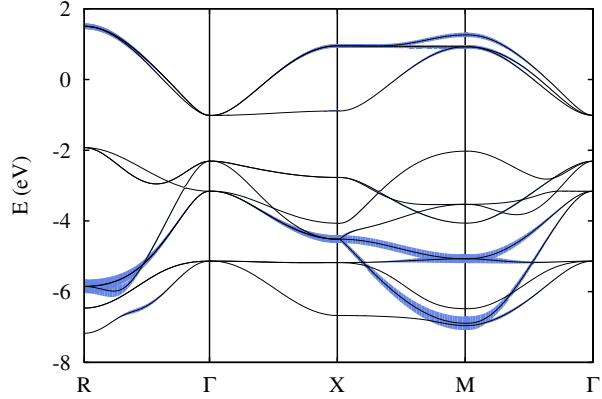}
\caption{Difference between the amplitudes of the orthonormalized $t_{2g}$ projectors in the PLO(0) and PLO(A) methods for V $t_{2g}$ as well as the O $2p$ bands in the projection. $15$ eV$\cdot\delta_n(\mathbf{k})$ (see equation \ref{delta_nk}) is shown as the width of the corresponding bands.
}
\label{diff_proj}
\end{figure}

The $e_g$ states of vanadium are mostly located above the $t_{2g}$ bands as conventional ligand-field theory suggests \cite{cox}. The bands below the V $d$ bands, extending from -8 eV to -2 eV are predominantly of oxygen $2p$ character, but also show hybridization between O $2p$ and the V $d$ states (figure \ref{fat_bands}).\\
The whole series of transition metal oxides SrVO$_3$-CaVO$_3$-LaVO$_3$-YVO$_3$ ranging from the metal SrVO$_3$ to the insulator YVO$_3$ can be classified, following Zaanen \etal \cite{zsa}, as Mott-Hubbard systems, since the ligand $p$ bands are clearly separated from the transition metal $d$ states at the Fermi level.
Thus, the low energy physics of the material is mainly determined by the three V $t_{2g}$ bands around the Fermi level. This suggests to use only these bands for the construction of the effective low energy Hamiltonian for the DFT+DMFT calculations. Such a construction, however, will not give any information of the behaviour of the O $p$ states, yet this might be of crucial importance for the description of the physics of the material. An important example where this is the case is NiO, where the charge-transfer behaviour of the system cannot be described by taking only the Ni $d$ states into account \cite{kunes_nio1,kunes_nio2,nio_dc_paper}.

\subsubsection{Tight-Binding discussion of the 3 band case}

The tight-binding like Hamiltonian created via the PLO method from only the V $t_{2g}$ states is compared to other schemes for generating Wannier functions, the maximally localized Wannier functions (MLWF) as defined by Marzani and Vanderbilt \cite{marzani_vanderbilt} in the PAW and the FLAPW basis sets and the Nth order muffin-tin method (NMTO) \cite{andersen_nmto} in table \ref{hops}.
To compute the hopping matrix elements two steps are necessary. First, one obtains the Hamiltonian in the Wannier representation by application of the projection matrices to the Bloch Hamiltonian $H^{B}_{nn'}(\mathbf{k})=\varepsilon_{n}(\mathbf{k})\delta_{nn'}$ in the following way
\[H^{W}_{LL^\prime}(\mathbf{k})=\sum_{nn'}\bar{\mathcal P}^\mathcal C_{Ln}(\mathbf{k}) H^{B}_{nn'}(\mathbf{k})\left(\bar{\mathcal P}^\mathcal C_{L'n'}(\mathbf{k})\right)^\ast\]
Second, Fourier transformation of the k-dependent Wannier Hamiltonian $H^{W}_{LL'}(\mathbf{k})$
yields the on-site energies and hopping integrals
\[t_{LL'}^{\mathbf{R}\mathbf{R}'}=
	\frac{1}{N}\sum_{\mathbf{k}}^N\exp(i\mathbf{k}\cdot(\mathbf{R}-\mathbf{R}')) H^{W}_{LL'}(\mathbf{k}).\]
In above formula $L,L^\prime$ label the orbitals and 
$\mathbf{R},\mathbf{R}^\prime$ are lattice vectors.

\begin{figure}[t]
 \centering
  \includegraphics[width=0.5\textwidth]{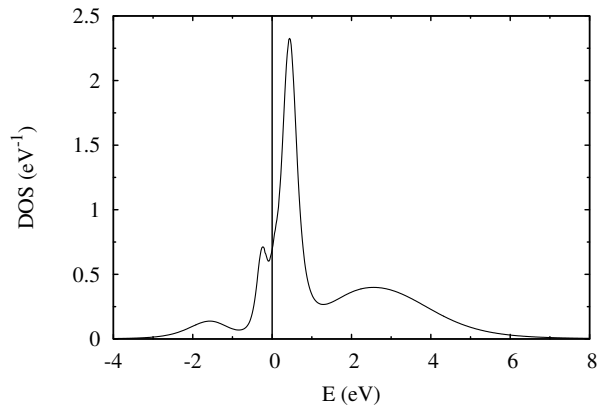}
\caption{Impurity spectral function obtained by DFT+DMFT(QMC) for $U=4$ eV at the inverse temperature $\beta=20~\mathrm{eV}^{-1}$. Only the three V $t_{2g}$ bands were used in the projection.}
\label{ed_u4_dos}
\end{figure}

The nearest neighbour hopping clearly dominates, which shows the short range of the bonding in SrVO$_3$. The three compared methods although very different in cost and conception yield virtually identical descriptions of the system. An explicit comparison of the PLO method with MLWFs generated by the WANNIER90 package \cite{wannier90} yields quantitative agreement with the present implementation in the present case. This underlines the validity of the our implementation. This is another instance of the known fact, that the first guess in the procedure of maximal localization in the MLWF scheme is usually already a very good one for transition metal oxides \cite{one_shot_wannier}. Since our implementation is such a \textit{first guess} procedure without any additional localization procedure the good agreement is not surprising. Additionally, the specifics of the system, like its high symmetry and the well separated block of $t_{2g}$ bands make it a rather elementary case.

\subsubsection{Comparison of different PLO methods}

Differences in the different PLO methods can become significant if the oxygen $2p$ bands are included. These states are essential in understanding the physics of many transition metal oxides. Their importance for the physics of the SrVO$_3$-CaVO$_3$-LaVO$_3$-YVO$_3$ series has been pointed out by Mossanek \etal \cite{mossanek2008}. The oxygen $2p$ bands are located below the V $t_{2g}$ block and hybridize considerably with them as shown in figure \ref{fat_bands}. The number of bands to be taken into account in the Wannier construction now is $12$. 
The resulting local densities of states (LDOS) for the effective three band model are shown in figure \ref{wannier_bands_12}. While the LDOS created by the PLO(A) and PLO(V) methods are virtually indistinguishable, the approximation PLO(0) creates a different LDOS. The distribution of the spectral weight is different in the PLO(0) method. In fact, the amplitudes of the projectors PLO(V) and PLO(A) are close to identical, while the PLO(0) projectors can differ significantly from the other two methods, leading to the difference in the resulting LDOS. The difference $\delta_n(\mathbf{k})$ between the projections PLO(0) $(\bar{\mathcal P}^\mathcal C_{Ln}(\mathbf{k})_{0})$ and PLO(A) $(\bar{\mathcal P}^\mathcal C_{Ln}(\mathbf{k})_{A})$ is shown pictorially in figure \ref{diff_proj} for every band $n$. Here the total difference between the amplitudes of the $t_{2g}$ projectors
\begin{equation}\delta_n(\mathbf{k})=\sum_{L}\Bigl\vert\left\vert\bar{\mathcal P}^\mathcal C_{Ln}(\mathbf{k})_{A}\right\vert-\left\vert\bar{\mathcal P}^\mathcal C_{Ln}(\mathbf{k})_{0}\right\vert\Bigr\vert^2
\label{delta_nk}
\end{equation}
\begin{figure}[t]
 \centering
  \includegraphics[width=0.5\textwidth]{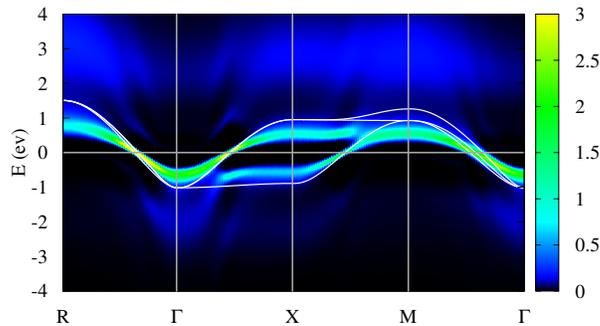}
\caption{Momentum resolved impurity spectral function obtained by DFT+DMFT(QMC) for $U=4$ eV. The three V $t_{2g}$ bands were used in the projection. As a comparison the DFT(LDA) band structure of the V $t_{2g}$ Bloch states is shown.}
\label{qmc_u4_spectral}
\end{figure}
has been computed. For better visibility $15$ eV$\cdot\delta_n(\mathbf{k})$ is shown as the width of the corresponding bands. On the same scale the difference between PLO(A) and PLO(V) would be by far smaller than the linewidth used for drawing the LDA bands in figure \ref{diff_proj}. The values attained in this case are $\delta_n(\mathbf{k})<5.5\cdot 10^{-5}$. This figure gives a qualitative picture of where the higher order terms of the expansion equation (\ref{eqn:LK_proj_full}) give significant contributions. This is the case outside of an energy window extending from $-0.5$ eV to $-2.5$ eV where no significant difference between PLO(0) and PLO(A) is observed. The quite large differences outside this window stem from the fact that the first order projector is constructed as the atomic eigenfunction and thus the projected bands from it overweight the lower lying bands leading to an underestimation of the spectral weight at and above the Fermi level. This leads to a higher occupancy of the effective impurity in the PLO(0) approximation as compared to the other methods. The PLO(A) and PLO(V) methods yield impurity occupancies of 0.7 electrons per orbital (including spin degeneracy), while the PLO(0) yields a filling of 0.85. This stems from the fact, that the impurity level is much lower in PLO(0), because the lower lying bands are overweighted. In fact, the impurity level (static crystal field) lies at about $-1.1$ eV when using the PLO(0) method, whereas it it is centered at approximately $-0.6$ eV for the other methods. A difference between the PLO(A) and PLO(V) constructions is that in the former a specific radial function $\ket{\phi_{\nu=0}}$ is chosen for the projection, while in the latter an averaging over radial dependencies takes place by the inclusion of general overlaps $\braket{\phi_{\nu}|\phi_{\nu'}}$. This did not make any noteworthy difference here or in other systems we have considered.

\subsubsection{DFT+DMFT calculations}

Let us now turn to the results obtained by DFT+DMFT using the projection scheme explained above. We have performed calculations using the HF-QMC solver for the material. We obtain momentum resolved spectral functions and are thus also able to compare our results with recent ARPES studies \cite{yoshida2005,yoshida2010}. For the calculations including only the $t_{2g}$ bands we have used the on-site interaction $U=4$ eV and for the calculations including also the O $p$ bands $U=6$ eV was used. In all cases $J=0.65$ eV was applied. This set of parameters was chosen to be able to compare with earlier studies of the material that used the same values \cite{lechermann_wannier,Amadon08}. Calculations were performed at inverse temperature $\beta=20~eV^{-1}$ using $200$ time slices and $10^6$ Monte-Carlo sweeps. Spectral functions were obtained from the measured Green function via the maximum entropy method \cite{maxent}. 

 \begin{figure}[t]
  \centering
  \includegraphics[width=0.49\textwidth]{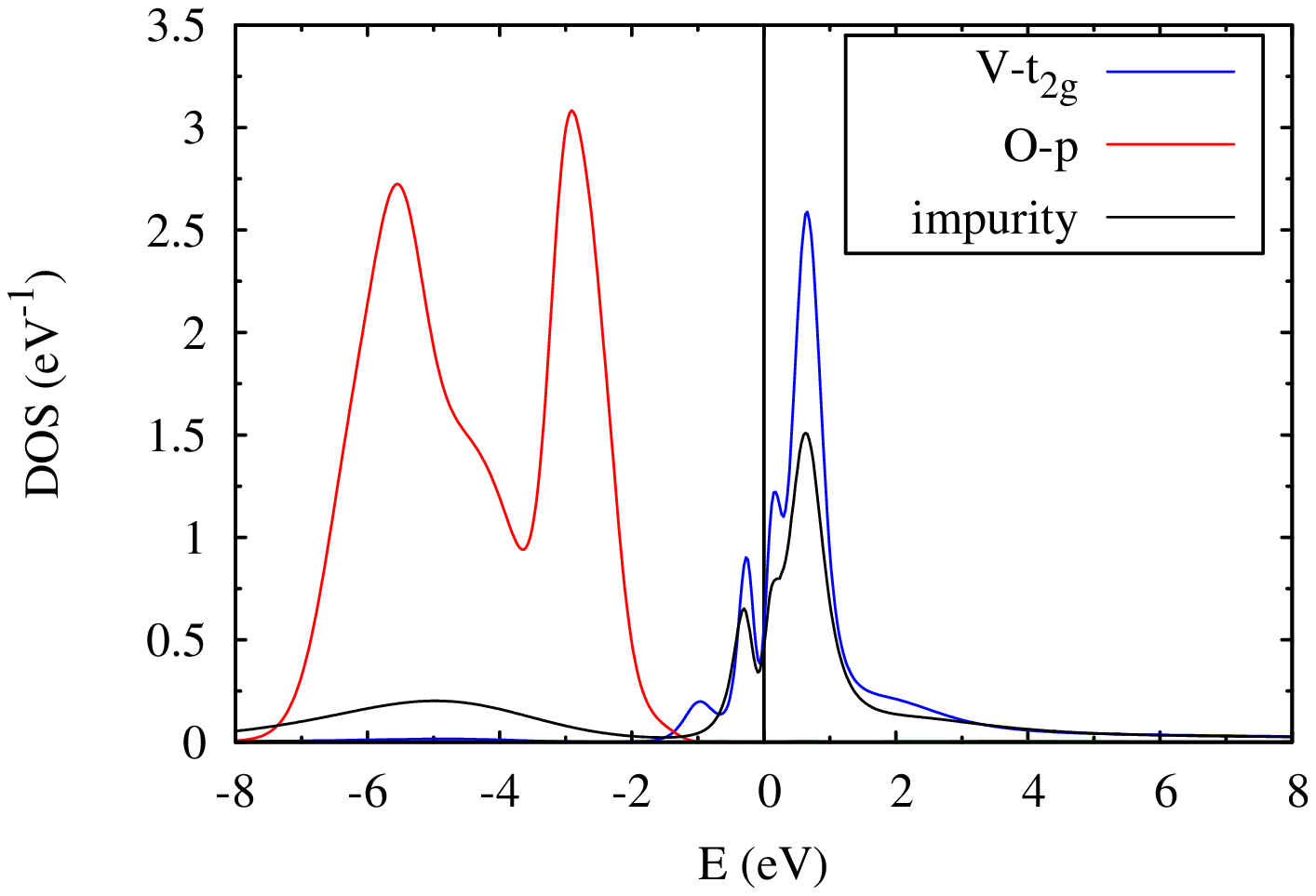}
  \hspace*{0pt}
   \includegraphics[width=0.49\textwidth]{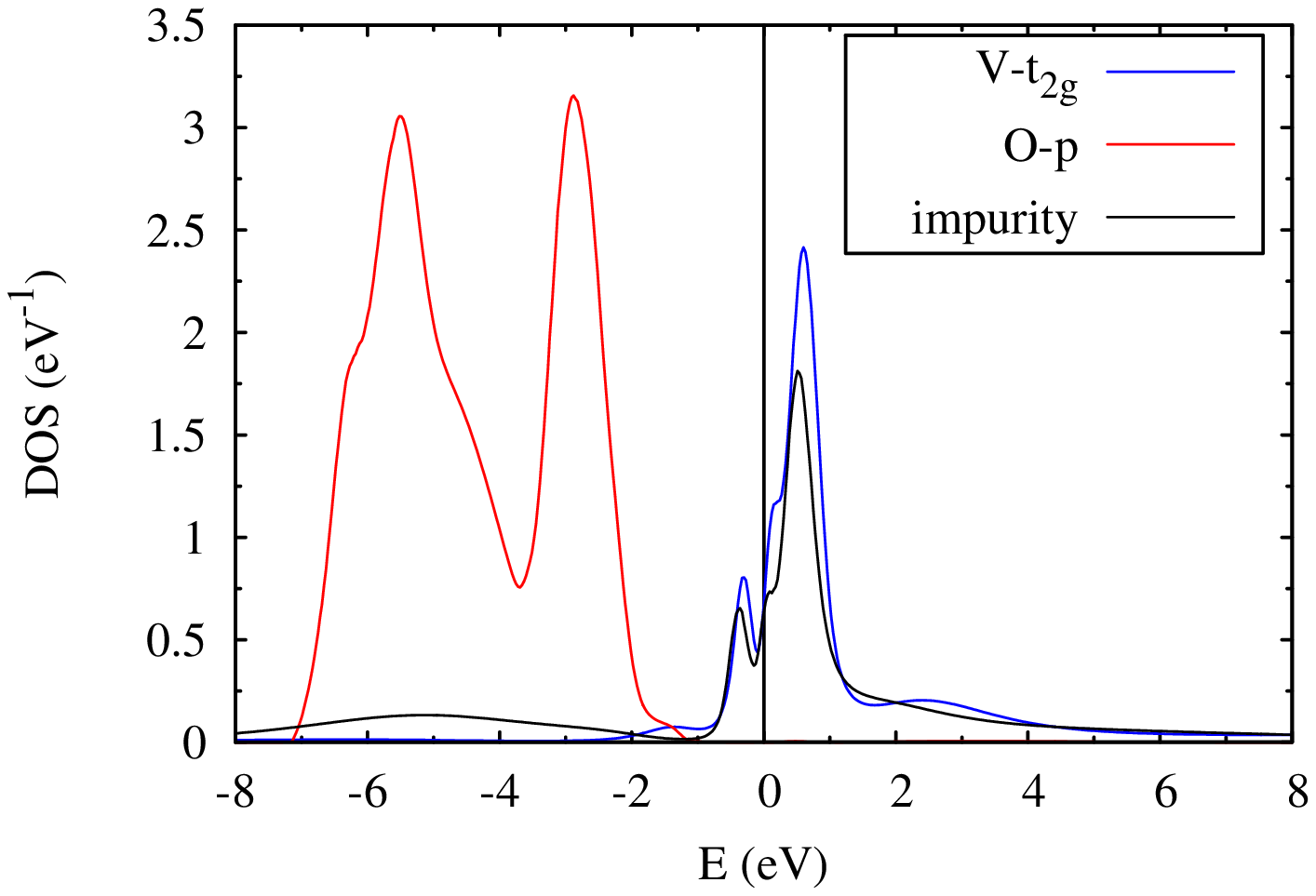}
 \caption{Orbitally resolved spectral function obtained by DFT+DMFT(QMC) and PLO(0) (left) and PLO(V) (right) for $U=6$ eV. The V $t_{2g}$ and O $2p$ states were used in the projection.}
 \label{ed_u6_dos}
 \end{figure}

\begin{figure}[t]
  \centering
   \includegraphics[width=0.5\textwidth]{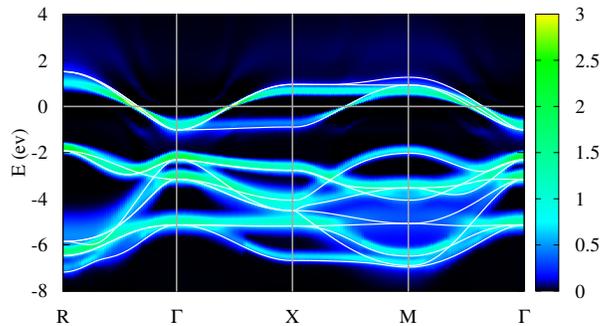}
 \caption{Momentum resolved impurity spectral function obtained by DFT+DMFT(QMC) and PLO(V) for $U=6$ eV. The V $t_{2g}$ and the O $2p$ states were used in the projection (12 bands). Again the LDA band structure of the V $t_{2g}$ and O $2p$ Bloch states is shown for comparison.}
 \label{k-resolved_spectrum}
 \end{figure}

First we examine results obtained by the minimal 3 band model obtained only from the three V $t_{2g}$ bands. As in earlier studies \cite{lechermann_wannier,Amadon08,nekrasov2006,Aichhorn2009} we obtain the local impurity spectral function for the V $t_{2g}$ states. The typical three-peak structure is apparent in the spectral function in figure \ref{ed_u4_dos}, with upper and lower Hubbard bands at $\sim3$ eV above and at $\sim2$ eV below the Fermi level, respectively. 
The data agree very well with other studies \cite{lechermann_wannier,Amadon08,nekrasov2006,Aichhorn2009}. We also estimated the mass renormalization 
\[Z=\left[1-\left.\frac{\partial }{\partial\omega}\mathrm{Re}\Sigma(\omega)\right\vert_{\omega=0}\right]^{-1}\] 
from the imaginary part of the self energy at the first Matsubara frequency to be $Z=0.61$, which means, that the mass enhancement is $m^\ast/m\sim1.65$ in accordance with ARPES estimates of $m^\ast/m\sim1.8\pm0.2$ \cite{yoshida2005}.
We also computed the momentum-resolved spectral function from the QMC data
\[A_i(\mathbf{k},\omega)=-\frac{1}{\pi}\mathrm{Im}\left(\omega+\mu-\varepsilon_i(\mathbf{k})-\Sigma_i(\omega)\right)^{-1},\]
which is shown in figure \ref{qmc_u4_spectral} together with the three V $t_{2g}$ DFT(LDA) bands as a comparison.
The Hubbard bands are clearly recognizable as non-coherent features around $-2$ eV and in the range of $2$ eV to $4$ eV above the Fermi level. The agreement with ARPES energy dispersions is generally quite good, especially the bottom of the quasiparticle band is found at $-0.6$ eV in contrast to the LDA value of $-1$ eV \cite{yoshida2005}. The overall width of the quasiparticle bands around the Fermi level is reduced to about $1.5$ eV, similarly to earlier reported data \cite{nekrasov2006}.\\
Since the inclusion of the oxygen $2p$ states into the model is important we also show results for that situation. The Wannier functions created in this case will be more localized than in the $t_{2g}$ only model, because they are now composed of a larger number of Bloch functions. The effective impurity problem is now constructed encompassing also the $p$ bands and thus also the residual $d$ spectral weight contained in them, see figure \ref{fat_bands}. The impurity spectral function will now also have some spectral weight inside the oxygen block. The spectral functions obtained via two different PLO methods while using QMC as impurity solver within DMFT are shown in figure \ref{ed_u6_dos} for PLO(0) and PLO(V), respectively. In this case the parameter values $U=6$ eV and $J=0.65$ eV were applied.
The impurity clearly shows the spectral weight inside the oxygen block which is stronger for the PLO(0) for reasons outlined above. The quasi-particle renormalization is $Z=0.62$ (from PLO(V), from PLO(0) it is slightly higher at $Z=0.66$)
in this case, which is in agreement with the previous $t_{2g}$-only estimate. Along with the impurity spectral function figure \ref{ed_u6_dos} also shows the spectral function decomposed by the respective Bloch bands. These can be obtained by applying the inverse of the projection matrices and thus upfolding the impurity back to the Bloch basis. 
The assignment of a certain band to a certain group, say O $p$ is performed using the dominant character of the band, as in the DFT analysis. The lowest lying nine bands are thus labeled as oxygen bands, the following three bands are labeled V $t_{2g}$. 
The differences between the PLO(0) and PLO(V) methods carry over into the DFT+DMFT description. The effective impurity shows considerably more spectral weight inside the oxygen block around $-5$ eV for the PLO(0) consistent with the LDA LDOS. While the PLO(A) and PLO(V) yield occupancies of 0.66 per orbital one finds 0.81 with PLO(0). The differently distributed spectral weight also gives rise to different densities of states at the Fermi level. For the PLO(V) case it is $\sim0.62~\mathrm{eV}^{-1}$ while it is reduced to $\sim0.56~\mathrm{eV}^{-1}$ for PLO(0). Furthermore, the quasi-particle peak is considerably reduced in PLO(0) at the expense of the enhanced spectral weight at lower energies. The upper and lower Hubbard bands also appear shifted in the direction of the Fermi level by some $0.5$ eV. These effects show, that the PLO(0) approximation does not capture the whole spectral weight of the $t_{2g}$ states around the Fermi level correctly.
The momentum-resolved spectral function, calculated using 12 bands in the Hamiltonian, is shown in figure \ref{k-resolved_spectrum}. The $t_{2g}$ bands around the Fermi level are in close agreement with the ones obtained in the $t_{2g}$-only case, showing a considerable renormalization as compared to the LDA bands, as before in accord with experimental studies. The upper Hubbard band is visible at 3 eV above the Fermi level, while the lower Hubbard band is hidden inside the oxygen bands starting at about 2 eV below the Fermi energy. The oxygen bands appear slightly broadened and shifted with respect to the $t_{2g}$ bands, but essentially unchanged as compared to the LDA bands.  

The inclusion of oxygen states gives rise to the so-called double counting problem of DFT+DMFT, or more generally of any DFT++ method which includes correlated and uncorrelated states.

To avoid a double counting in the energy a term $H_{\rm dc}=\mu_{dc}\sum_{m,\sigma} n_{m,\sigma}$, with the double counting potential $\mu_{dc}$ and orbital index $m$, has to be subtracted from the Hamiltonian, because the local Coulomb interaction is already contained in DFT in an averaged manner. The double-counting amounts to a shift of the chemical potential of the correlated orbitals and is an issue, since there is no direct microscopic or diagrammatic link between the DFT and the model Hamiltonian part. In the DFT+DMFT implementation that we used a constraint on the electronic charge to define a double counting correction. The occupancies of the correlated subspace computed from the local non-interacting Green function and the interacting impurity Green function are required to be identical, which can be stated in the following form
\begin{equation}
\mathrm{Tr}~G^{imp}_{mm^\prime}(\beta)\stackrel{!}{=}\mathrm{Tr}~G^{0,loc}_{mm^\prime}(\beta).
\label{crit1}
\end{equation}
The double counting potential $\mu_{dc}$ is iteratively adjusted to fulfill equation (\ref{crit1}). The most obvious effect of a change in the double counting is a shift of the oxygen states versus the V $d$ states. The effect of the double counting is not so dramatic for the case at hand (as shown in Ref. \cite{Amadon08}), yet it influences quantitative results. This problem can be resolved by explicitly calculating the interaction between the $d$ and the $p$ states ($U_{pd}$) or in methods like GW+DMFT \cite{gw_dmft}, where the double counting is explicitly available.

\subsection{Mott transition in Ca$_{2-x}$Sr$_x$RuO$_4$}

As a second application we present the Mott transition in Ca$_{2-x}$Sr$_x$RuO$_4$ at small Sr contents ($x\leq 0.1$).
This system is an example of a strongly distorted perovskite crystal, where the local symmetry is reduced. We will show that the PLO methodology we presented can also be applied to systems with lower symmetry and that we can obtain accurate results concerning the coupled structural and electronic metal-insulator transition in the system. The system has been studied during recent years,
since it shows a variety of interesting structural, electronic and magnetic 
phases \cite{braden, friedt}.
Especially the discovery of unconventional superconductivity in the pure Sr 
compound $(x=2)$ by \cite{Maeno_nature} has caused interest in the correlated 
electronic structure of these ruthenates. Substitution of Sr$^{2+}$ with the smaller, but isoelectronic, Ca$^{2+}$ results in 
structural distortions of the cubic ($I4_1/mmm$) crystal structure of 
Sr$_{2}$RuO$_4$. These distortions reduce the crystal symmetry to the $Pbca$ 
symmetry group and transform the system into an antiferromagnetic Mott insulator 
at low temperatures in the doping range (0 $\leq$ $x$ $<$ 0.2) \cite{friedt}.

\begin{figure}[t]
 \centering
  \includegraphics[width=0.2\textwidth]{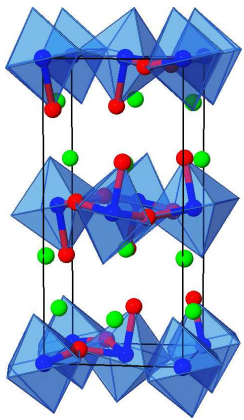}
\caption{Crystal structure of $S-Pbca$ Ca$_2$RuO$_4$. Only atoms contained in the unit cell are shown: Ru (blue), O (red) and Ca (green).
The octahedron surrounding every Ru atom has O atoms on each of its corners. The Ru layers are separated by rock salt-like layers of Calcium and Oxygen.}
\label{crystal_struct}
\end{figure}

The low energy physics of the material series is governed mainly by the Ru-$4d(t_{2g})$ bands close to the Fermi level \cite{Woods} that are occupied by 4 electrons.
The tetragonal symmetry eventually leads to an 
additional ``2D'' crystal-field splitting within the $t_{2g}$ bands, breaking their 
degeneracy \cite{Fang_Nagaosa_Terakura}. We will refer to this additional splitting as \textit{the} crystal-field (CF) splitting. 
The crystal structure of Ca$_2$RuO$_4$ in the low temperature $S-Pbca$ phase is shown in figure \ref{crystal_struct}.
The structural distortions can be classified as rotations of the RuO$_6$ 
octahedra around two different axes and an additional compression or elongation of 
the octahedra along certain of their symmetry axes as described in Ref. \cite{friedt}. The effects on the electronic structure can be understood in terms of hybridization of the Ru($4d$) orbitals with the neighbouring O($2p$) 
orbitals as has been investigated in Refs. \cite{Fang_Terakura, Fang_Nagaosa_Terakura}.
The coupled structural and electronic metal-insulator transition occurring in the material \cite{alexander_1999} has been a matter of controversy.
On increasing the on-site Coulomb interaction $U$ Anisimov \etal \cite{osmt_Anisimov} found an orbital 
selective Mott transition in the $t_{2g}$ bands. 
A different mechanism for the metal-insulator transition was proposed by Liebsch and Ishida \cite{liebsch_ishida} who investigated 
the transition with respect to the Coulomb interaction $U$ and the CF splitting and found a common metal-insulator transition in all $t_{2g}$ orbitals. 
In both works local quantum correlations were taken into account correctly. The effects of crystal distortions were, however, neglected completely or only incorporated using model parameters, like the CF splitting.
Since both structural modifications \textit{and} local dynamical correlations govern the physics of the system, it is 
imperative to include the structural distortions in a true \textit{ab initio} manner. 

\subsubsection{DFT}
\begin{figure}[t]
  \centering
   \includegraphics[width=0.5\textwidth]{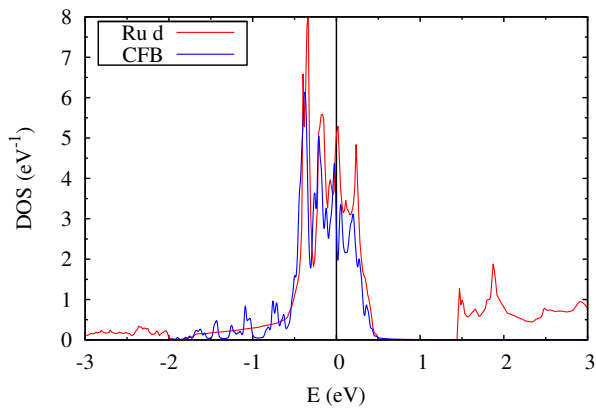}
 \caption{Effective impurity local density of states in the crystal field basis (CFB) of the $S-Pbca$ structure of Ca$_2$RuO$_4$ for the effective three band model (blue) compared to the full Ru $d$ density of states (red).}
 \label{csro_ldos}
 \end{figure}

For the ruthenate system three Wannier orbitals corresponding to $t_{2g}$ symmetry 
are used. The construction of those orbitals is complicated by the symmetry-reducing
structural distortions in Ca$_{2-x}$Sr$_{x}$RuO$_4$. 
The PAW-projectors are defined with respect to an intrinsic (global) cubic coordinate system which coincides with the crystallographic axes in the present case. 
The crystal distortions in the system, especially the rotational distortions, locally modify the orientation of the orbitals inside the Ca/Sr cube surrounding the RuO$_6$ octahedra. This leads to the fact, that the global symmetry character $t_{2g}$ vs. $e_g$ for the Ru($4d$) orbitals is not resolved satisfactory and we obtain a mixture of global $t_{2g}$ and $e_g$ at the Fermi level.
\begin{figure}[t]
\includegraphics[width=0.5\textwidth]{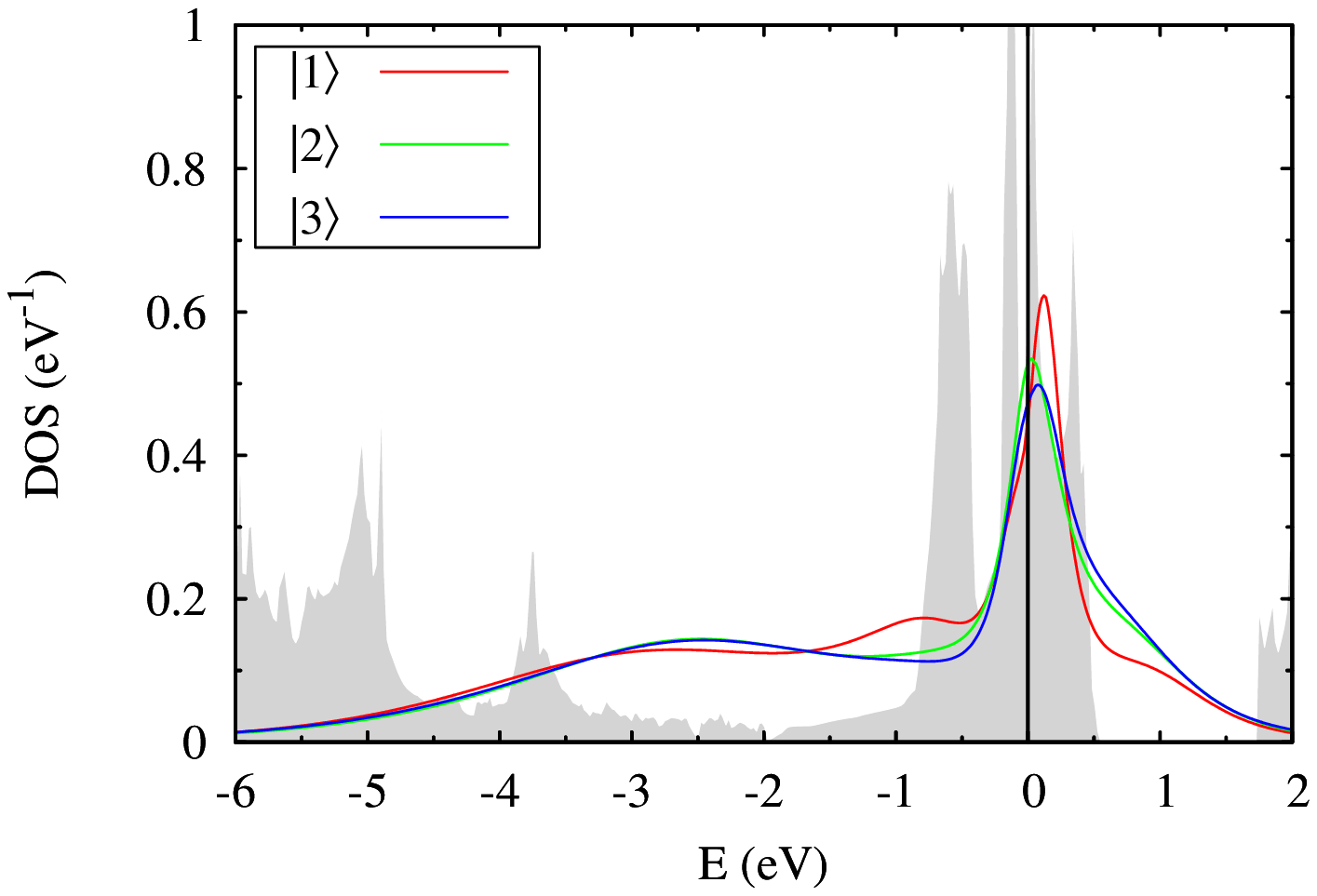}
\hspace*{20pt}
\includegraphics[width=0.5\textwidth]{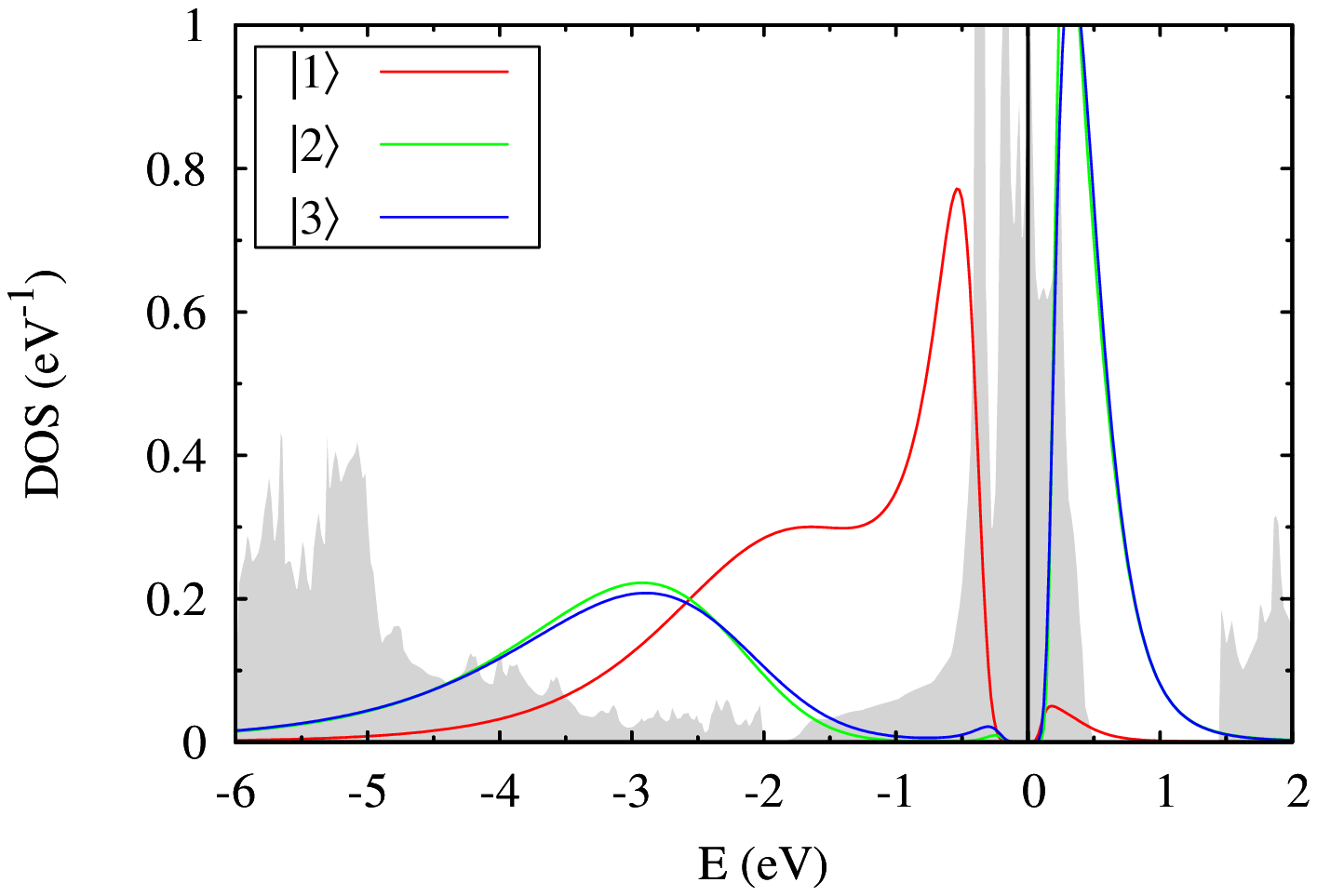}
\caption{DMFT spectral functions for the metallic structure at $\beta=20~\mathrm{eV}^{-1}$ (left) and for the insulating one $\beta=30~\mathrm{eV}^{-1}$ (right) showing the evolution towards the insulating state for $x=0.0$. The Ru($4d$) density of states from DFT(PBE) is shown in gray.}
\label{fig 1}
\end{figure}
Since it is known from previous 
calculations~\cite{Woods,Fang_Terakura,Fang_Nagaosa_Terakura} that the electronic 
structure of Ca$_{2-x}$Sr$_{x}$RuO$_4$ around the Fermi level is governed by bands deriving
from the Ru($t_{2g}$) states, local rotations of the orbitals have to be considered in order
to be able to obtain the correct orbitally resolved density of states and 
correct projectors in the PLO framework. In this context \textit{local} means that 
the orbitals are rotated on each Ru site in the crystal cell.
The angular parts of the five $4d$ orbitals are represented by real spherical harmonics $K_{lm}$ 
(also called cubic harmonics) corresponding to the angular momentum $l=2$. These 
functions are derived from the real and imaginary parts of the complex spherical 
harmonics $Y_{lm}$ (see e.g. Ref. \cite{coppens}).
Their transformation properties under rotations are therefore related to the well 
known transformation properties of the $Y_{lm}$ \cite{rose, edmonds} and were given in e.g. Ref. \cite{coppens}. 
The Ru($d$) projectors in the energy interval considered 
($\sim$-2.5 eV to $\sim$+0.5 eV) should show only few $e_g$ character
since the cubic CF splitting moves the $e_g$ states to energies of $\sim$ 1.5eV above
the Fermi level. Therefore in a first step the contribution of the $e_g$ orbitals to
the projection was minimized. This was achieved by a rotation of the underlying coordinate system. 
The Wannier Hamiltonian in the rotated basis was subsequently diagonalized on-site. 
The latter yields the so called crystal-field basis. The orbitals thus obtained are linear combinations of the rotated $t_{2g}$ orbitals 
only. They are named in ascending order by their energy as 
$\left(|1\rangle\ |2\rangle\ |3\rangle \right)$. 
In the case of Ca$_{2-x}$Sr$_{x}$RuO$_4$, the crystal-field basis retains most of 
the character of the underlying rotated $t_{2g}$ orbitals in the investigated doping
range. For $x=0.0$ in the insulating ($S-Pbca$) structure, e.g., it reads
\[
		\left(\begin{array}{c}
		|1\rangle\\
		|2\rangle\\
		|3\rangle
		\end{array}\right)=		
		\left(
		\begin{array}{rrr}
		0.997 &  -0.046 & -0.056 \\
		0.055 & 0.978 &0.201 \\
		0.045 & -0.203 &0.978
		\end{array}	
		\right)
		 \left(\begin{array}{c}
		|xy\rangle\\
		|xz\rangle\\
		|yz\rangle
		\end{array}\right).
\]
The LDOS obtained from the effective three band model in the CF basis
is shown in comparison to the \textit{total} Ru-$4d$ DOS (the VASP code package and the Perdew-Burke-Ernzerhof (PBE) gradient corrected functional \cite{pbe} were used) in figure(\ref{csro_ldos}). The agreement concerning the bandwidth and general features of the DOS is good, discrepancies stem mostly from the $e_g$ states, that have been explicitly projected out from the CF basis.  The effective three band impurity is correctly occupied by 4 electrons.

Four different crystal structures (obtained via high resolution powder diffraction by \cite{friedt}) have been considered, to ensure that we can picture the metal-insulator 
transition (MIT) both with respect to the temperature $T$ as well as to Sr content 
$x$, i.e. $x=0$ at 180 K (insulating) and at 400 K (metallic), $x=0.1$ at 10 K 
(insulating) and at 300 K (metallic) \cite{Friedt_Phd}. The temperatures here refer to those at which the 
structural measurements were performed in Refs. \cite{friedt} and 
\cite{Friedt_Phd}. We will refer to the structures as low temperature (LT) 
and high temperature (HT) for every $x$. Disorder effects emerging from Ca $\to$ Sr substitution have not been taken into account; we always calculated Ca$_2$RuO$_4$ in different crystal structures.

\begin{table}[ht]
 \caption{Crystal Field Splittings and root-mean-square hopping integrals (see text) for all structures considered. \label{cf splittings}}
 \begin{indented}
 \item[]\begin{tabular}{@{\extracolsep{\fill}} c c c c c c}
 \br
 &\multicolumn{2}{c}{$x=0.0$}&\multicolumn{2}{c}{$x=0.1$}\\	
 \mr
structure & LT & HT&LT&HT\\
$\Delta$ [meV] &$295$ & $95$ &  $236$ & $90$\\
$t_{rms}[meV]$ & $152$ & $201$ &  $167$ & $205$\\
\br
\end{tabular}
 \end{indented}
 \end{table}
 \begin{figure}[t]
 \includegraphics[width=0.5\textwidth]{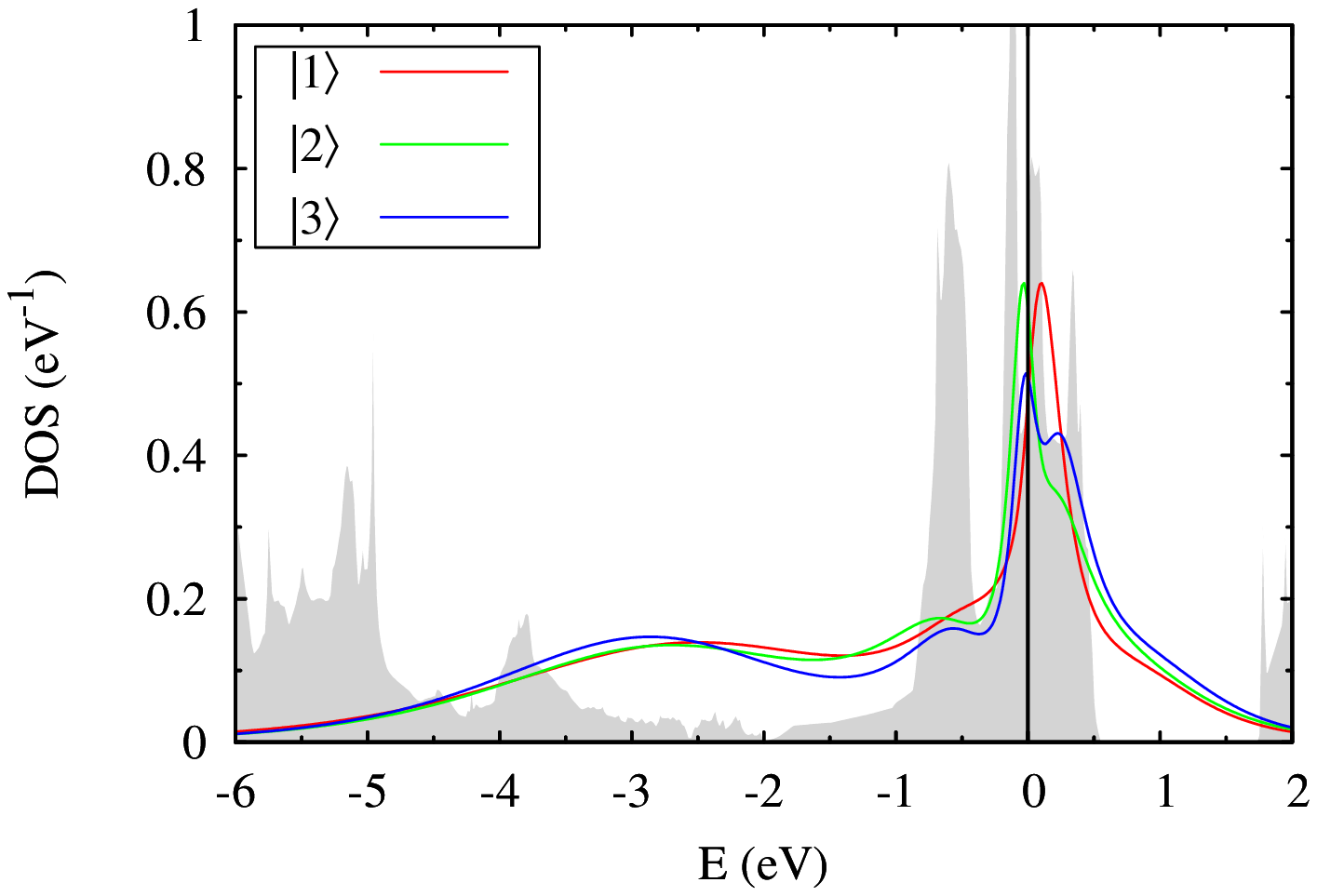}
\hspace*{20pt}
  \includegraphics[width=0.5\textwidth]{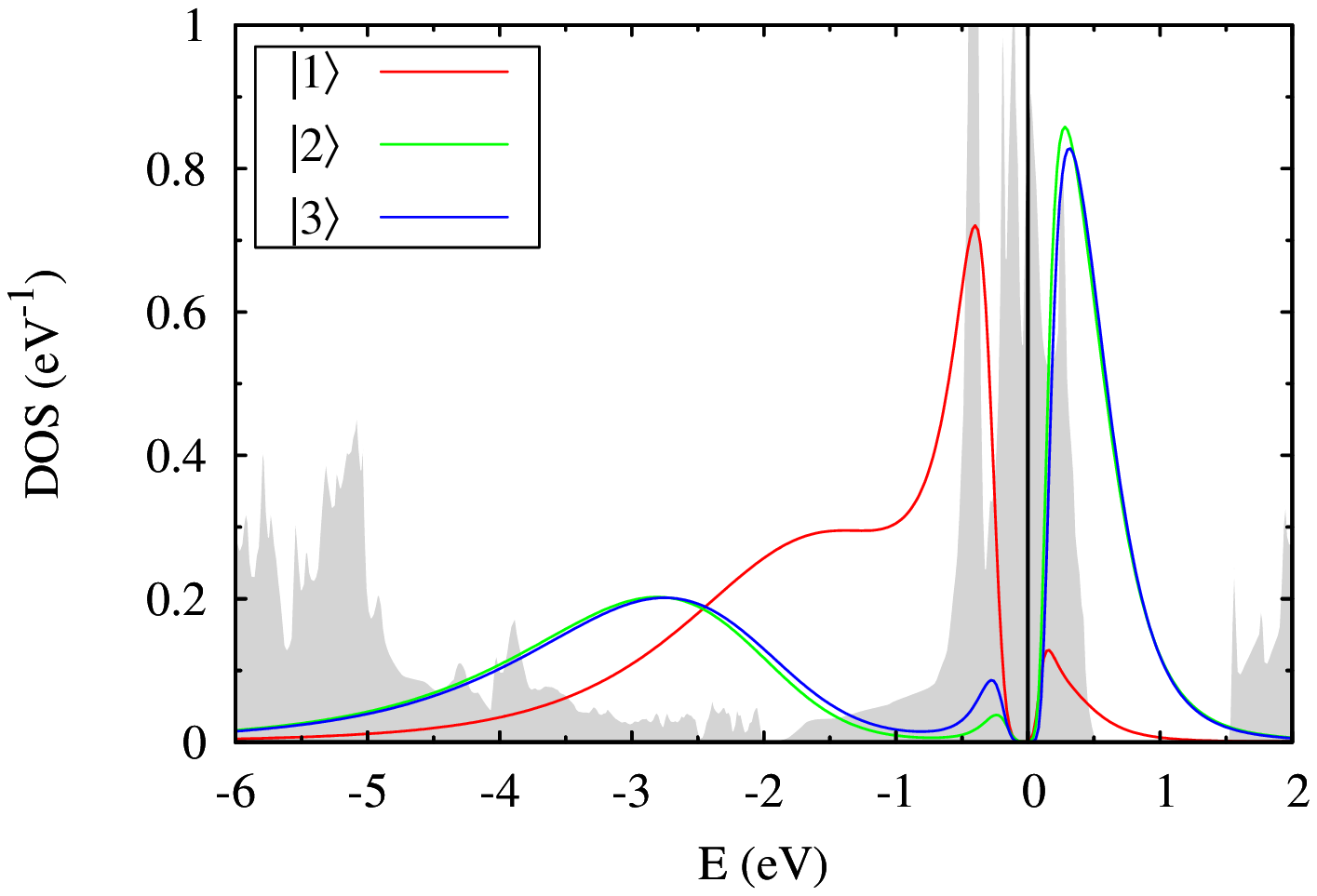}
\caption{DMFT spectral functions for the metallic structure at $\beta=20~\mathrm{eV}^{-1}$ (left) and for the insulating one $\beta=30~\mathrm{eV}^{-1}$ (right) showing the evolution towards the insulating state for $x=0.1$. The Ru($4d$) density of states from DFT(PBE) is shown in gray.}
\label{fig 2}
\end{figure}

It is insightful to first consider the tight-binding picture obtained from the Wannier hamiltonian.
The crystal field splitting 
$\Delta=\varepsilon_{2,3}-\varepsilon_1$ for each of the structures considered is 
given in Tab. \ref{cf splittings}. The values for 
$\Delta_{21}=\varepsilon_{2}-\varepsilon_1$ and 
$\Delta_{31}=\varepsilon_{3}-\varepsilon_1$ are very similar, since the orbitals 
$|2\rangle$ and $|3\rangle$ are very close in energy. For clarity only one value is 
given in Tab. \ref{cf splittings}.
It is apparent that the CF splitting is strongly enhanced for the low temperature 
phases at $x=0$ and $x=0.1$. This is also reflected in the DFT(PBE) occupancies for those
structures, namely $(n_1,n_2,n_3)=(0.70,0.65,0.65)$ and $(0.69,0.66,0.65)$ for the
HT and  $(0.79,0.62,0.60)$ and $(0.75,0.63,0.62)$ for the LT structures at $x=0$ 
and $x=0.1$ respectively. The $|1\rangle$ orbital is considerably lowered in energy,
its population increases via $\left(|2\rangle,|3\rangle\right)\to|1\rangle$ charge
transfer upon change from the HT to the LT phase.
The nearest neighbour hopping integrals given in table \ref{cf splittings} also show a
clear picture. We define the root-mean square hopping
$t^2_{rms}=\frac{1}{N_o\cdot N_{n}}\sum_{(o,n)} t^2$ as a relative measure of the 
kinetic energy\footnote{this is in the 
spirit of Ref. \cite{pavarini_review}} . The sum runs over all nearest neighbours and all orbitals, the 
normalization factor is given by the product of the number of orbitals $N_o$ and 
the number of nearest neighbours $N_n$. The values for the LT structures at $x=0$ 
and $x=0.1$ are significantly smaller than those for the HT structures. Our DFT 
results hence show an increasing orbital polarization and electron localization as 
$x$ decreases and also on the transition from HT to LT. This is in accord with 
previous DFT calculations \cite{Fang_Terakura, Fang_Nagaosa_Terakura}.

\subsubsection{DFT+DMFT calculations}

Let us now turn to the DFT+DMFT results. We parametrize the on-site Coulomb interaction matrix by $U_{mm}=U$, $U_{mm^\prime}=U-2J$, and $J_{m\neq m^\prime}=J$ \cite{PhysRevB.56.12909}, with $U=3.1$ eV and $J=0.7$ eV 
as obtained from constrained DFT calculations by Pchelkina \etal \cite{pchelkina} 
for Sr$_{2}$RuO$_4$. Since we employ the HF-QMC solver to the resulting 3 orbital model, only density-density type interaction terms are included. The neglected terms have important effects on the quantitative description of the transition. The results presented therefore show an approximate description of the system, see the discussion in Ref. \cite{ca2ruo4_prl}.
The computations were performed down to a temperature 
of $\beta=20~\frac{1}{eV}$ ($\sim$580 K) for the HT structures and $\beta=30~\frac{1}{eV}$ ($\sim$387 K) for the LT ones. We focus on the paramagnetic metal-insulator transition exclusively. 
The spectral functions, shown in figures 
\ref{fig 1} and \ref{fig 2} with the Ru($4d$) DFT DOS as a reference, were obtained 
from the imaginary time Green functions by analytic continuation via the Maximum 
Entropy Method \cite{maxent}. 
A small insulating gap begins to form at $x=0.1$ and becomes clearly visible at $x=0$ in the LT phase. 
The behaviour of orbital $|1\rangle$ is qualitatively different from the orbitals 
$|2\rangle,|3\rangle$. This is also reflected in the occupancies within DFT+DMFT. 
At $\beta=30~\frac{1}{eV}$ for $x=0.1$ and $x=0.0$ they read $(0.95,0.52,0.53)$ and $(0.98,0.51,0.51)$, 
respectively. Thus, the local correlations introduced via the Hubbard $U$ strongly enhance the occupation of the $|1\rangle$ orbital driving it towards
a band-insulating state. The other orbitals approach half filling and undergo a
standard Mott-transition. In the high temperature structures the introduction of local correlations into the system does not alter the results 
obtained by DFT qualitatively. The system remains in a correlated metallic state.
The DMFT occupation numbers for the HT structures at $x=0$ and $x=0.1$ consequently 
remain almost exactly at their DFT values.

The calculated spectral functions are in good agreement with photoemission data for Ca$_2$RuO$_4$ \cite{mizokawa_2001}. 
The quasiparticle peak in the high-temperature phase, the Hubbard bands at $-0.5$ eV,
and at $-3$ eV can be identified with corresponding features in
the experimental spectra. The peak at about $-1.5$ eV binding energy in the experimental data reported in Ref. \cite{mizokawa_2001}
is most probably a contribution of O $2p$ states, which are not included in our calculations. The small gap of about $0.2$ eV
is in agreement with electrical resistivity\cite{alexander_1999,nakatsuji_2000_1,nakatsuji_2000_2}
and optical conductivity \cite{puchkov_1998} measurements for Ca$_2$RuO$_4$. The orbital order in the $S-Pbca$ phase and its disappearance in the $L-Pbca$ phase is also in line with the evolution of x-ray absorption spectra (XAS) across the metal-insulator transition for $x=0$ and $x=0.09$ \cite{mizokawa_2001,mizokawa_2004}.
The observed mechanism of the metal-insulator transition is in accord with the mechanism suggested by \cite{liebsch_ishida}, 
but not to be reconciled with the orbital selective scenario. 
An orbital selective Mott transition would require successive Mott transitions in 
orbitals $|2\rangle,|3\rangle$ and $|1\rangle$ driven by the different widths of 
the subbands. The bandwidth however, which was assumed to be the driving force 
behind the orbital selective transition is of secondary importance in the system.

\section{Summary and Conclusions}

We have presented a general interface between a projector augmented wave based all-electron DFT method and many body methods based on Wannier functions obtained from a projection on local orbitals. Different schemes to obtain projection matrices from PAW calculations have been explored and explicitly compared to other schemes, like Nth order muffin-tin orbitals or maximally localized Wannier functions for the cubic perovskite SrVO$_3$.

We find that care has to be taken to correctly represent all the spectral weight of the correlated subspace in the construction of the projected local orbitals. The simplest scheme, taking into account only the first PAW projector can be a bad approximation and leads to an incomplete description of the system. Only constructions taking into account higher PAW projectors are capable of a correct description of correlation physics on the LDA as well as on the LDA++ level.
Our k-resolved spectral functions show dispersive quasiparticle features around the Fermi energy resembling renormalized LDA bands as well as incoherent features that can be identified as Hubbard bands. Mass renormalization factors are consistent with previous experimental \cite{yoshida2005} as well as theoretical studies \cite{pavarini_andersen,nekrasov2006}.
Subsequently we have also studied the distorted perovskite system Ca$_{2-x}$Sr$_{x}$RuO$_4$ for $x\leq0.1$, where a local symmetry adaption of the PLO projection matrices is necessary. In accordance with experimental \cite{alexander_1999,nakatsuji_2000_1,nakatsuji_2000_2,mizokawa_2001,mizokawa_2004,puchkov_1998} and recent theoretical studies \cite{liebsch_ishida, ca2ruo4_prl}, we identify a mechanism for the common structural and a metal-insulator phase transition: The structural changes in the system enhance the crystal field splitting leading to a Mott transition in some orbitals and to a band insulating behaviour in other orbitals. 

Our DFT++ implementation is very flexible and allows for applications ranging from the bulk systems, discussed here, to magnetic nanostructures or isolated correlated impurities \cite{co_on_cu,co_on_graphene}.

\ack

We thank M. Marsman and G. Kresse for providing us the VASP to WANNIER90 interface. Support from SFB 668 is acknowledged.

\end{document}